\documentclass[10pt]{iopart}
\usepackage{epsfig}
\usepackage{iopams}
\eqnobysec

\begin{document}

\title{Strategies for observing extreme mass ratio inspirals}

\author{Steve Drasco}

\address{Jet Propulsion Laboratory, California Institute of Technology \\ Pasadena, California 91109}

\date{\today}

\begin{abstract}
I review the status of research, conducted by a variety of independent groups,  
aimed at the eventual observation of Extreme Mass Ratio Inspirals (EMRIs) with 
gravitational wave detectors.  EMRIs are binary systems in which one of the objects
is much more massive than the other, and which are in a state of dynamical evolution that is 
dominated by the effects of gravitational radiation.  Although these systems are highly
relativistic, with the smaller object moving relative to the larger at nearly light-speed, 
they are well described by perturbative calculations which exploit the mass ratio as a natural
small parameter.  I review the use of such approximations to generate waveforms
needed by data analysis algorithms for observation.  I also briefly
review the status of developing the data analysis algorithms themselves.  Although this article is 
almost entirely a review of previous work, it includes (as an appendix) a new analytical estimate for the 
time over which the influence of radiation on the binary itself is observationally negligible.
\end{abstract}


\maketitle

\section{EMRIs and IMRIs}\label{EMRIs and IMRIs}

This article is intended to be a brief review of a variety of works which share a common ultimate goal: the 
observation of Extreme Mass Ratio Inspirals (EMRIs) with gravitational wave detectors.  I 
assume that the reader is familiar enough with the subject that she believes such work to be worthwhile, 
but not so familiar that she has kept up with the most recent progress (astrophysical motivation can be 
found in the introduction of almost any of the more recent references below).
This article is intended to report on such progress.  It is not intended to be an in-depth review, such as 
Refs.~\cite{glampedakis review, poisson livrev, sasaki tagoshi}.  With that said, I now very briefly describe EMRIs.

An EMRI is made up of a non-spinning test mass $\mu$ orbiting a black hole with mass $M$ and angular momentum 
of magnitude $aM < M^2$ (I use units where $G = c = 1$).  
I will use Boyer-Lindquist coordinates $(t,r,\theta,\phi)$ to describe 
the background geometry of the black hole.
The test mass perturbs the spacetime metric $g_{\alpha\beta}$ so that it deviates from the Kerr 
metric $g^{\rm{Kerr}}_{\alpha\beta}(M,a)$
\begin{equation}
g_{\alpha\beta} = g^{\rm{Kerr}}_{\alpha\beta}(M,a) + h_{\alpha\beta} \;,
\end{equation}
where the perturbation is of order the mass ratio $h_{\alpha\beta} \sim \mu/M$.
Assuming that the test mass is restricted to a region near the boundary beyond which no bound Kerr geodesics exist
(that is, near the radius $r_{\rm{ISCO}}$ of the innermost stable circular orbit, $r \gtrsim r_{\rm{ISCO}}$) 
the mass ratio $\mu/M$ determines the nature of the system's evolution.  
That is, it determines whether the test mass quickly plunges into the hole, or instead moves 
in some sort of persistent ``orbital motion''.  This can be seen from a crude scaling argument.  
Suppose that the test mass moves on some sort of bound orbit about the hole.  A distant 
observer would measure the energy of the test mass to be $E \sim \mu$ and the period of the orbital 
motion to be $T_{\rm{orb}} \sim M$.  The power of the observed radiation would 
be\footnote{This relation can of course be derived, but the sceptic might recall some other familiar example 
of waves where power is proportional to the square of the wave's amplitude.  Here, 
the wave's amplitude is the metric perturbation $h_{\alpha\beta} \propto \mu/M$.} $dE/dt \sim (\mu/M)^2$. 
The timescale over which the orbital energy changes $T_{\rm{rad}}$ would be the ratio of the orbital energy to the 
radiative power $T_{\rm{rad}} \sim M^2/\mu$.  The supposition that the test mass orbits the hole is then self-consistent when
\begin{equation}
\frac{ T_{\rm{orb}} }{ T_{\rm{rad}} }\sim \frac{\mu}{M}  \ll 1 \;.
\label{self cons}
\end{equation}
The EMRIs with frequencies $f = 1/T_{\rm{orb}}$ in the band of LISA-like detectors will have masses in the ranges 
$10^5\lesssim M/M_\odot\lesssim 10^7$ and $1\lesssim \mu/M_\odot\lesssim 10^2$.  To a good approximation, these 
systems satisfy the condition for orbit-like motion (\ref{self cons}).  See Sec.~III B of Ref.~\cite{hughes 2000} for a more rigorous estimate 
of when to expect orbit-like motion.

A close relative of EMRIs are asymmetric binaries with frequencies in the band of LIGO-like detectors.  These 
intermediate mass ratio inspirals (IMRIs) have mass ranges $10^2\lesssim M/M_\odot\lesssim 10^3$ and 
$1\lesssim \mu/M_\odot\lesssim 10$.  IMRIs of course satisfy the orbit-like motion condition to a lesser extent than EMRIs, 
but it may still be useful to search for them with techniques that were designed for EMRIs.   Searching 
for IMRIs with LIGO-like detectors has caught the attention of astronomers since it amounts to a search for intermediate 
mass black holes.  It is attractive to those working with EMRIs because of the similar physics, and 
because searches can be performed immediately.  LIGO's S5 science run could likely observe IMRIs out to a modest 
distance of roughly 10 to 60 Mpc.  Though this range is unlikely to yield detections, advanced LIGO will be sensitive to 
IMRIs out to a much more promising distance of about 0.2--0.9 Gpc \cite{brown et al}.

\section{The EMR in EMRI}

It's useful to first become familiar with the ``EMR'' part of EMRIs before discussing the ``I''.  That is, it's useful to understand 
the geodesic orbits of test particles bound to  black holes before considering the effects of radiation.  This is motivated in part 
because EMRIs will spend the majority of their lifetimes in the regime of orbit-like motion, where Eq.~(\ref{self cons}) holds.  
Also, recent developments in the description of these orbits have become powerful tools which should find applications beyond EMRIs.

Compared to Newtonian orbits, strong field black hole orbits have unfamiliar properties.  Orbits close to the hole have three 
distinct orbital frequencies.  This is because, unlike weak-field orbits which are planar, the orbit is confined within a toroidal 
region with three degrees of freedom.  One way to define the boundaries of this torus is to choose values for the three 
constants of geodesic motion: energy $E$, axial angular momentum $L$, and Carter constant $Q$ (the Kerr-analog of the 
magnitude of the non-axial angular momentum).  The following coordinate-based definition of the orbital torus is often more intuitive.

As the orbit rotates azimuthaly about the spin axis of the hole, it bounces between two radii $r_{\min} < r < r_{\max}$.
The radial boundaries are often defined in terms of an eccentricity $e$ and a semilatus rectum $p$,
both of which conform to their Newtonian definition in the weak field
\begin{equation}
\frac{r_{\min}}{M} = \frac{p}{ 1 + e }\;,
~~~~
\frac{r_{\max}}{M} = \frac{p}{ 1 - e} \;.
\end{equation}
Some authors omit the mass $M$ here, such that their $p$ would have dimensions of length.
The polar motion of the orbit will also be bounded by some minimum angle $\theta_{\min} \le \theta$.
Since the black hole is symmetric under reflection about its equatorial plane, the other polar boundary 
is redundant $\theta \le \pi - \theta_{\min}$.
Alternatively, one can define these boundaries with an inclination angle\footnote{This definition of $\iota$ is different from the 
one found in Ref.~\cite{drasco hughes 2006} and elsewhere.  In Ref.~\cite{drasco hughes 2006}, what I am currently calling 
$\iota$ was defined as $\theta_{\rm{inc}}$.} $\iota$
\begin{equation}
\iota + ({\rm{sgn~}} L) \theta_{\min} = \frac{\pi}{2} \;.
\end{equation}
The ${\rm{sgn~}} L$ term is in place so that $\iota$ varies continuously 
from 0 to $180^\circ$ as orbits go from prograde to retrograde.  For weak-field orbits, $\iota$ is the angle between the 
orbital and equatorial planes.  In the strong field however, orbits are not at all planar.  In that case $\iota$ is an indicator 
of the ``polar thickness'' of the orbital torus.  It indicates only the boundary within which the orbit bounces in and out of 
the equatorial plane.

\begin{figure}
\begin{center}
\epsfig{file=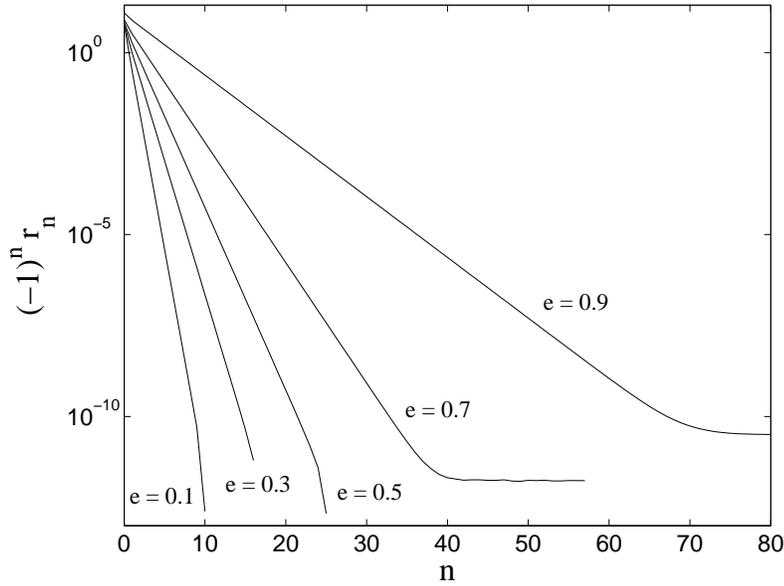,angle=0,width=10.5cm}
\caption{Fourier series coefficients for $r(\lambda)$ in the case of a black hole spin of $a = 0.9M$, and geodesics 
with $p=6$, $\iota = 40^\circ$.  The analogous series for $\theta(\lambda)$, $\phi(\lambda)$, and $t(\lambda)$ 
converge similarly.  The two curves which level out ($e = 0.7$ and $e = 0.9$) appear to have exhausted the 
precision capabilities of the numerical method used to produce this plot.}
\label{f:rn}
\end{center}
\end{figure}

Neither the azimuthal ($\phi$), polar ($\theta$), or the radial ($r$) motions are periodic functions of  the time $t$ measured 
by a distant observer's clock, or even of proper time $\tau$.  The radial and polar motion are however periodic functions of 
Mino-time $\lambda$ (defined by $d\tau/d\lambda = r^2 + a^2 \cos^2 \theta$) \cite{mino 2003}.  If 
the orbit begins from $r = r_{\min}$ and $\theta = \theta_{\min}$, when $\lambda = 0$, then
\begin{equation} \label{series}
r(\lambda) = r_0 + 2\sum_{n = 1}^\infty r_n \cos (n \Upsilon_r \lambda) \;,~~~~
\theta(\lambda) = \frac{\pi}{2} + 2\sum_{k = 1}^\infty \theta_k \cos (k \Upsilon_\theta \lambda) \;, 
\end{equation}
where $r_n$, $\theta_k$, and $\Upsilon_{r,\theta}$ are constants.  The functions $t(\lambda)$ 
and $\phi(\lambda)$ have
similar harmonic decompositions except that they each (i) increase linearly with $\lambda$, and (ii) contain harmonics of 
\emph{both} $\Upsilon_r$ and $\Upsilon_\theta$.
These decompositions have proved powerful tools for two reasons.  First, as Fig.~\ref{f:rn} demonstrates, the series 
(\ref{series}) generally converge rapidly. 
Even for orbits with large eccentricity or inclination,  one need only compute a small number of coefficients in the Fourier 
series in order to evaluate the orbit with great accuracy at arbitrary times.  Second, and somewhat surprisingly, 
it turns out that by exploiting these decompositions, one can show that in the frequency domain associated with 
coordinate time $t$, most functions of these orbits have a discrete spectrum made up of frequencies \cite{drasco hughes 2004}
\begin{equation} \label{fmkn}
f_{mkn} = m f_\phi + k f_\theta + n f_r \;.
\end{equation}
Here $m$, $k$, and $n$ are integers, while $f_r$, $f_\theta$, and $f_\phi$ are fixed frequencies 
determined by the boundaries of the orbital torus.  For example, if a test particle moves on a fixed geodesic, a distant observer 
will find that the resulting gravitational waves oscillate only at frequencies $f_{mkn}$.

In the presence of radiation, returning the ``I'' to EMRI, waveforms observed by distant observers would have a sliding-comb type 
of frequency-spectrum, a discrete spectrum which over time slides around on the frequency axis.  Since these waveforms 
will likely be weak compared to detector noise, models for the evolution of these spectra will be needed in order to enable 
some form of matched filtering.  The remainder of this article will review EMRI-related work which falls into one of two 
categories: (i) the calculation of waveforms, and (ii) the development of data analysis algorithms.  Since my own research falls 
into the first category, I will spend more time on the discussion of waveform calculations than on data analysis development.  
Of course, my bias should  not be taken as an indicator of relative importance.

\section{Waveforms}

Efforts to compute EMRI waveforms can roughly be grouped into three categories:
(i) Capra waveforms, (ii) Teukolsky waveforms, and (iii) kludge waveforms. 
I define the categories as follows (these definitions are described in the following subsections): 
If a waveform calculation is based on solving the MiSaTaQuWa equations \cite{misata,quwa}, or some higher order version of them, the result
is a Capra waveform.  If a waveform calculation is based on solving the Teukolsky equation \cite{teukolsky 1972, teukolsky 1973, ryan}, the result is
a Teukolsky waveform.  If a waveform calculation is based on a variety of formalisms, some of which may
even make conflicting assumptions, the result is a kludge waveform.
I have listed these categories roughly in order of increasing availability and decreasing accuracy. 
However, the ultimate categorization factor will be taken to be the method of calculation rather than 
either its accuracy or availability.  I now describe these waveforms, and the status of efforts to compute them

\subsection{Capra waveforms}

The name Capra comes from the Hollywood film director Frank Capra (1897-1991)
whose ranch (which now belongs to Caltech) served as the location for the first of
a now-annual series of ``Capra meetings''.    The ultimate goal of
these meetings is to compute a class of waveforms that incorporates the leading-order (in $\mu/M$) effects 
of both the radiative and conservative parts of the gravitational self-force.
The self-force acts on the test mass and is induced by that same object's perturbation of the background spacetime
(it is a force in the sense that, instead of interpreting the motion of the test mass as a geodesic of some perturbed version 
of the background spacetime, it can be interpreted as being \emph{forced} to deviate from geodesics of the background.).
The radiative part of the self-force produces radiation at the black hole's horizon and at large radial distances.  The conservative 
part does not produce such radiation, but nonetheless influences the world line of the small object, and the corresponding 
waveform \cite{pound poisson nickel}.  Perhaps the crown jewels of efforts toward Capra waveforms are the equations of 
motion for a test particle under the influence of the first-order parts of the self-force in an arbitrary background spacetime.  
These are the so-called MiSaTaQuWa equations \cite{misata,quwa}.  I categorize any waveform as a Capra waveform if 
its calculation was based on solving the MiSaTaQuWa equations, or some higher order version of them. Capra waveforms 
can be thought of as ``holy grail waveforms''.  In terms of potential accuracy, they are the most ambitious.  
They are the only waveforms capable of being computed to an accuracy of order $(\mu/M)^2$. 
Realizing this potential will require computing all of the first-order parts of the self-force, and at least some of its second-order 
parts \cite{rosenthal, hinderer flanagan}. 

The holy grail characterization of Capra waveforms applies both to their accuracy and availability. 
To date, no calculation of Capra waveforms exists 
(with the exception of analog problems for scalar or electromagnetic fields \cite{pound poisson nickel}).
Perhaps the most advanced effort toward Capra waveforms presented at the 2005 Capra meeting \cite{capra 2005}
is the work by Barack and Lousto \cite{barack lousto}.
They developed a numeric code that can compute a waveform for a test mass on an arbitrary worldline in a Schwarzschild 
background.  The code evolves the metric perturbation in the Lorenz gauge, so that the input worldline can
be given by the solution of the mode-sum representation of the MiSaTaQuWa equations \cite{barack ori 2000, barack et al, barack ori 2003}
(known only in the Lorenz gauge, but for any orbit in either Schwarzschild or Kerr backgrounds). 
Instead of such a worldline, a circular Schwarzschild geodesic is used in Ref.~\cite{barack lousto}. 
They successfully compared the power radiated to infinity with results from calculations based on Teukolsky waveforms 
(discussed below).  Incorporating the solution of the mode-sum representation of the MiSaTaQuWa equations into the code from 
Ref.~\cite{barack lousto} would produce the first of any subcategory of Capra waveforms.

While progress toward Capra waveforms is steady, much work remains.  Second-order versions of the MiSaTaQuWa 
equations and their solutions must be derived.  Then, codes similar to the one used in Ref.~\cite{barack lousto} must 
be developed before the full potential of Capra waveforms will be realized.  More details on this subject can be found 
in Poisson's short manifesto \cite{poisson 2004} or in his comprehensive treatise \cite{poisson livrev}.

\subsection{Teukolsky waveforms}

In 1972, Teukolsky derived a powerful equation for describing first-order radiative perturbations of black holes 
\cite{teukolsky 1972, teukolsky 1973, ryan}.  It applies to cases where the hole is perturbed by a scalar, neutrino, 
electromagnetic, or gravitational field. Although his is a partial differential equation, Teukolsky showed that it can be 
separated into a system of ordinary differential equations (ODEs).  Solving the Teukolsky equation numerically is 
therefore far simpler than traditional numerical relativity \cite{baumgarte shapiro}.  I categorize any waveform as a 
Teukolsky waveform if its calculation was based on solving the Teukolsky equation.  Since solutions of the Teukolsky 
equation describe only radiative first-order (in $\mu/M$ for the case of EMRIs) effects, Teukolsky waveforms have less potential 
accuracy than Capra waveforms which, at least in principle, can also describe second-order and conservative effects.  

The calculation of Teukolsky waveforms has been something of an industry since the early 1990's (see table I of 
Ref.~\cite{drasco hughes 2006}).  These calculations exploit the condition that the test mass moves in an orbit-like 
fashion, so that Eq.~(\ref{self cons}) holds.  The source term in the Teukolsky equation is taken to be a point-particle on 
a bound geodesic of the background spacetime.   Since the equation's solution contains all the radiative information, it 
describes the waveform produced by the particle's motion, and also the effect of that radiation on the orbit.  By iteration, 
the orbit evolves from a geodesic to an inspiral while the waveform evolves from only a ``snapshot'' of 
what an observer would see over a short time, to the entire EMRI waveform.  This strategy has only recently been applied 
to generic black hole orbits with both eccentricity $e \ne 0$ and inclination $\iota \ne 0^\circ,180^\circ$.
The radiation snapshots produced by these orbits have been computed numerically \cite{hdff, drasco hughes 2006, drasco hughes data}, but the 
iteration needed to produce the full EMRI waveform is still under development.  I now sketch a few of the details of these new developments.

As is described in Sec.~\ref{EMRIs and IMRIs}, many functions of generic black hole orbits have a discrete frequency-spectrum at 
frequencies $f_{mkn}$, given by Eq.~(\ref{fmkn}).  The radiation produced by a test mass moving on that orbit is one such function.  
A distant observer would see that radiation as the following waveform
\begin{equation} \label{snapshot}
\fl
h_+ - i h_\times = -\frac{2}{r}
\sum_{l = 2}^\infty
\sum_{m = -l}^l
\sum_{k = -\infty}^\infty
\sum_{n = -\infty}^\infty
 \frac{Z^{\rm{H}}_{lmkn}}{\omega_{mkn}^{2}}
                   S_{lmkn}(\theta) e^{-i\omega_{mkn}(t-r)+im\phi + i\chi_{lmkn}}\;.
\end{equation}
Here $h_+$ and $h_\times$ are the two independent components of the metric perturbation, 
$\omega_{mkn} = 2\pi f_{mkn}$, and both the coefficients $Z^{\rm{H}}_{lmkn}$ and the functions $S_{lmkn}(\theta)$ are found by solving 
the ODEs that separate out from the Teukolsky equation 
\footnote{
These quantities are defined in Sec.~III of Ref.~\cite{drasco hughes 2006}.  In the case of perturbations to a black hole's spacetime geometry, 
the leading order correction to the (otherwise vanishing) Weyl curvature scalar $\psi_4$, a quantity that completely describes the radiation in 
the perturbed spacetime, satisfies the Teukolsky equation.  
In the case of perturbations from a bound test particle, $\psi_4$ can be simply projected onto a basis of angular functions $S_{lmkn}(\theta)$ and 
radial functions $R_{lmkn}(r)$ which both depend on the orbit's fundamental frequencies $\omega_{mkn}$.
The angular functions satisfy an ODE which, apart its dependence on $\omega_{mkn}$, is homogeneous.
The radial functions satisfy an inhomogeneous ODE such that
$R_{lmkn}(r \to \infty) = Z^{\rm{H}}_{lmkn} f^{\rm{H}}(\omega_{mkn},r)$ and $R_{lmkn}(r \to r_+) = Z^{\infty}_{lmkn}f^{\infty}(\omega_{mkn},r)$, 
where the event horizon is located at $r = r_+$. The explicit (analytically known) definitions of the functions $f^{\rm{H},\infty}(\omega_{mkn},r)$ are 
beyond the scope of this review.
}.  The phase constants $\chi_{lmkn}$ in Eq.~(\ref{snapshot}) are determined by the initial 
position of the test mass \footnote{The exact dependence is shown in Eq.~(8.29) of Ref.~\cite{drasco flanagan hughes}.}, and can be 
set to zero when considering a fixed orbit. Similar expressions for the radiative changes in the orbiting particle's energy $E$, 
axial angular momentum $L$, and Carter constant $Q$ have been derived \cite{hdff, drasco flanagan hughes, sago et al 2005, sago et al 2006}.  
For example, the average change in the orbital energy is given by
\begin{equation}\label{Edot}
\fl
\left< \frac{dE}{dt} \right> = 
\sum_{l = 2}^\infty
\sum_{m = -l}^l
\sum_{k = -\infty}^\infty
\sum_{n = -\infty}^\infty
\frac{1}{4\pi\omega_{mkn}^2}
\left ( \left|Z^{\rm{H}}_{lmkn}\right|^2 + \alpha_{lmkn}\left|Z^\infty_{lmkn}\right|^2 \right) \;,
\end{equation}
where $ \alpha_{lmkn}$ are simple constants that are analytically known \cite{drasco hughes 2006}, and 
$Z^\infty_{lmkn}$ are similar to $Z^{\rm{H}}_{lmkn}$.   
The geometry of the orbit dictates which terms are significant in Eqs.~(\ref{snapshot}) and (\ref{Edot}).  If the orbit 
is both circular and equatorial, all of the terms with $k \ne 0$ and $n \ne 0$ vanish.  Similarly, for orbits with eccentricity 
above about $0.3$ and inclination below about $60^\circ$, one captures better than half of the radiation by keeping only 
the terms with $k = 0$ \cite{drasco hughes 2006}.

Summing over only one of the terms in the parentheses of Eq.~(\ref{Edot}) gives either the power radiated at large radial distances, 
or into the horizon. One normally would arrive at this formula by first deriving each sum independently, and then 
enforcing global energy conservation \cite{teukolsky press}.  This derivation has a major drawback.  It is not generalizable to 
the evolution of the Carter constant.  Alternatively, one can derive the same result as follows. First, solve the Teukolsky 
equation (this gives the radiation field, or equivalently $Z^\infty_{lmkn}$ and $Z^{\rm{H}}_{lmkn}$).  Second, express the 
radiative self-force in terms of the radiation field.  Third, express the average rate of change in the orbital constant of interest 
in terms of the radiative self-force.  Galt'sov showed that, under general circumstances, this gives the traditional result 
\cite{teukolsky press} when applied to both energy and angular momentum \cite{galtsov}.  At the time however, the validity 
of using only the radiative self-force was not known.  Mino has since proved this to be valid \cite{mino 2003}, and the 
method has now been used to describe the evolution of the Carter constant in terms of the same quantities used for 
energy and angular momentum (namely $Z^\infty_{lmkn}$ and $Z^{\rm{H}}_{lmkn}$) \cite{hdff, drasco flanagan hughes, sago et al 2005, sago et al 2006}.  
Although the final equation for $\left<dQ/dt\right>$ in Refs.~\cite{hdff, drasco flanagan hughes} appear different from the more 
concise result in Refs.~\cite{sago et al 2005, sago et al 2006}, their equivalence has since been shown \cite{sago drasco}.

It has recently been demonstrated that, for an analogous case of an electric charge moving on a Newtonian orbit, one 
must account for  the evolution of $\chi_{lmkn}$ in Eq.~(\ref{snapshot}) in order to compute the full EMRI 
waveform to leading order in the charge to mass ratio \cite{pound poisson nickel}.  Also, for the case considered 
in Ref.~\cite{pound poisson nickel}, the evolution of $\chi_{lmkn}$ was completely determined by the conservative 
self-force alone.  Solutions to the Teukolsky equation contain only radiative information, so if this result carries over 
to the strong field regime relevant to EMRI observations, it will severely decrease the usefulness of both Teukolsky 
waveforms and kludge waveforms (described below).  
It should be emphasized however that the analysis in Ref.~\cite{pound poisson nickel} applies in the limit where gravity is Newtonian, and that some of their 
findings are expected to be atypical of the strong field regime (see the last several paragraphs of Sec.~V in Ref.~\cite{pound poisson nickel}).
Even if their results apply partially to the strong field, so that in the strong field the evolution of $\chi_{lmkn}$ is observationally significant, but can be 
well approximated from the radiative self-force alone, the equations governing such an evolution have yet to be 
derived.  Though such equations should in principle be obtainable from the analysis in Appendix C of Ref.~\cite{mino 2004},    
the resolution to this problem remains a subject of current research \cite{hinderer flanagan}.

There is also a history of efforts to solve the Teukolsky equation without fully exploiting its separability---only the $\phi$-coordinate 
is separated such that the codes must solve a ``$2+1$'' dimensional partial differential equation.  
These ``time-domain'' calculations have yet to match the precision of their ``frequency-domain'' alternatives.
In the most generic time-domain calculation to date (equatorial or circular Kerr orbits \cite{khanna}) total fluxes of energy and 
angular momentum agreed with frequency-domain codes to within about 25\%, whereas independent frequency-domain 
codes often agree to as many as six digits \cite{drasco hughes 2006}.  However, recent improvements in time-domain methods, 
including the use of an adaptive mesh, are promising.  Such codes have so far been applied to  Schwarzschild \cite{sopuerta laguna}.  
There agreement with frequency-domain codes improved by about a factor of ten over the previous generation of time-domain codes, agreeing 
to within about 0.01\%.  See Ref.~\cite{sopuerta laguna} for a discussion of the possible advantages of time-domain methods.

\subsection{Kludge waveforms}

I categorize any waveform as a kludge waveform if its calculation was based on a collection of different formalisms, 
some of which may even make conflicting assumptions (i.e.~solving the flat-space quadrupole formula for a 
particle on a relativistic black hole orbit).  Kludge waveforms are in generally less accurate than Capra or Teukolsky waveforms.
However as more rigorous waveforms have become available, kludge waveforms have made good on their promise of capturing 
the dominant features of the more realistic waveforms \cite{gair glampedakis}.
Kluge waveforms are readily available, and can be computed very quickly.  This makes them the tool of choice when scoping out 
candidate data analysis techniques (discussed in the next  section).  Their flexible design allows one to experiment freely, although 
admittedly crudely, by adding and removing proposed physical effects which would be unimaginably difficult to incorporate into Capra 
or Teukolsky waveform calculations.  For example, kludge waveforms are already available for speculative non-Kerr background 
space-times that may be used in straw-man/null-experiment tests of general relativity \cite{collins hughes, glampedakis babak}. 
In principle, kludge waveforms can also include effects due to the conservative self-force \cite{gair glampedakis, babak et al}.
Kludge waveforms may be lacking in rigor and perfection, but they overflow with availability and adaptability. 

The majority of kludge waveform calculations are numerical.  I now describe the analytical exceptions.
As early as 1963, Peters and Mathews derived waveforms for eccentric Newtonian orbits  using the flat space quadrupole formula
\cite{peters mathews}.  Barack and Cutler \cite{barack cutler} stiched together sequences of those waveforms by using 
post-Newtonian formulas to evolve the constants of orbital motion.  This enabled them to account for relativistic effects including those 
due to the black hole's spin, precession of the perihelion, and Lense-Thirring precession.  The post-Newtonian formalism is generally 
invalid for EMRIs, since the test mass moves at nearly light-speed.  However, their waveforms were useful for making rough estimates 
of how well LISA could measure an EMRIs parameters \cite{barack cutler}.  These analytic kludge waveforms will also likely be used 
in the future for a LISA Mock Data Challenge, a project that will use simulated LISA data to test proposed data analysis techniques 
for a variety of sources (initial results will be announced at the 2006 GWDAW meeting).  Another analytic effort,
originally motivated as a tool for exploring convergence of post-Newtonian series and for improving waveforms for neutron star binaries, 
derives the post-Newtonian expansion to solutions of the Teukolsky equation \cite{sasaki tagoshi}.  This hybrid technique assumes both a small mass ratio and 
slow motion.  The most recent contribution to this effort applies to Kerr orbits which are both slightly inclined and slightly 
eccentric \cite{sago et al 2006}, and it is anticipated that the same calculation will soon be complete for arbitrary inclination \cite{sago private}.
Table \ref{hybrid vs Teuk} shows that the hybrid calculation of radiative fluxes of energy and angular momentum \cite{sago et al 2006}
compared well with pure Teukolsky calculations \cite{drasco hughes 2006}.
\begin{table}
\begin{tabular}{cc||cc} 
\hline
$e$ &
$\iota$ &
disagreement in radiative power &
disagreement in radiative torque \\
\hline
0.03 & $1.7^\circ$  & $0.019~20\%$ & $0.013~23\%$\\
0.03 & $3.4^\circ$  & $0.019~22\%$ & $0.013~22\%$\\
0.03 & $5.2^\circ$  & $0.019~26\%$ & $0.013~19\%$\\
\hline
0.06 & $1.7^\circ$  & $0.163~7~\%$ & $0.075~23\%$\\
0.06 & $3.4^\circ$  & $0.163~8~\%$ & $0.075~14\%$\\
0.06 & $5.2^\circ$  & $0.163~8~\%$ & $0.074~97\%$\\
\hline
0.09 & $1.7^\circ$  & $0.734~9~\%$ & $0.319~8~\%$\\
0.09 & $3.4^\circ$  & $0.734~9~\%$ & $0.319~4~\%$\\
0.09 & $5.2^\circ$  & $0.734~9~\%$ & $0.318~7~\%$\\
\hline
\end{tabular}
\caption{Comparison of numerical solutions to the Teukolsky equation \cite{drasco hughes 2006, drasco hughes data} with
analytical post-Newtonian expansions of those solutions \cite{sago et al 2006}.  
Here the black hole has spin $a = 0.9M$, and the orbits have semilatus rectum $p = 100$. 
The dependence of the disagreement on the orbital parameters is the expected result of the hybrid 
calculation's use of a truncated Taylor expansion in both eccentricity $e$ and inclination $\iota$.
\label{hybrid vs Teuk}}
\end{table}

The following is a rough outline of the procedure used to compute kluge waveforms numericaly:
First, evolve the orbital constants by integrating equations of the form
\begin{equation} \label{step 1}
\frac{d\mathcal{E}}{dt} = Y(\mathcal{E},a,M,\mu) \;,
\end{equation}
where $\mathcal{E} = (E, L, Q)$, or $\mathcal{E} = (e, p, \iota)$, to obtain $\mathcal{E}(t)$.
Since $(E,L,Q)$ can be translated exactly to $(e,p,\iota)$, and vice versa, the choice here is somewhat arbitrary.
These equations might be, say, analytical post-Newtonian approximations, or some numerical formula gleamed 
from a fit to numerical data from solving the Teukolsky equation.
Second, using these solutions, determine the inspiraling world line ${\bf x}(t)$ by integrating a system of 
geodesic equations of the form (e.g.~Carter's equations \cite{carter} if the background is Kerr)
\begin{equation} \label{step 2}
\frac{d{\bf x}}{dt} = {\bf V}[\mathcal{E}(t),a,M,\mu] \;.
\end{equation}
Third, obtain the waveform by solving an equation of the form
\begin{equation}\label{step 3}
h_{ij}({\bf x}_{\rm{field}} ,t) = W_{ij}({\bf x}_{\rm{field}},{\bf x}[t]) \;,
\end{equation}
where here the subscript ``field'' denotes the position of the observer.
Examples of such an equation might be the flat space quadruple, or quadruple plus octupole formulas.

\begin{figure}
\begin{center}
\epsfig{file=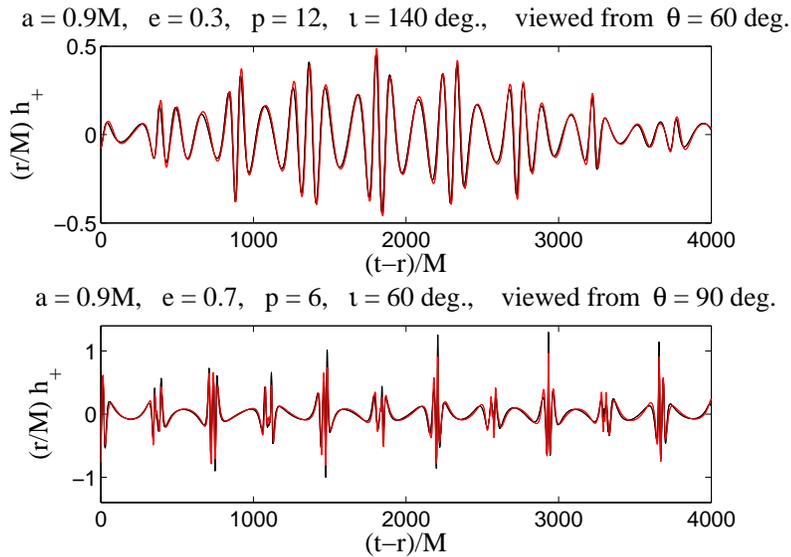,angle=0,width=10.5cm}
\caption{Comparison of Teukolsky \cite{drasco hughes 2006, drasco hughes data} (black) and numeric kludge \cite{babak et al} (red, or light gray) waveform snapshots for two 
different generic black hole orbits.  The parameters of the black hole and the orbits are shown in the titles above the plots. This level of agreement continues for essentially all time.  The short time range shown here was used so that
the reader can get a feel for the overlap by eye. The overlap estimates are 99\% for the top panel (over a time of 20,000 $M$) and 
97\% for the bottom panel (over a time of 15,000 $M$).}
\label{f:BFGGH}
\end{center}
\end{figure}

This numerical kludge procedure has been followed by several groups using a variety of different approximate 
formulas at each step \cite{ghk, gair glampedakis, babak et al}.  
The first step of this procedure, integrating the equation that evolves orbital constants (\ref{step 1}), 
has been executed by Gair and Glampedakis
for the case of generic black hole orbits \cite{gair glampedakis}.  In a separate calculation, 
for a fixed generic black hole orbit that does not inspiral,  
Babak et al \cite{babak et al} obtained snapshots of generic kludge waveforms by completing steps two and three 
[solving Eqs.~(\ref{step 2}) and (\ref{step 3}) with $\mathcal{E}$ held fixed].  
They found remarkable agreement when comparing with Teukolsky waveforms (see Fig.~\ref{f:BFGGH}).  
Estimates of the overlap between those kludge waveforms \cite{babak et al} and Teukolsky waveforms \cite{drasco hughes 2006, drasco hughes data} 
were typically around 98\%, for $p\gtrsim 6$.  This high level of agreement suggests that kludge waveforms may ultimately be used 
directly in EMRI detection algorithms, as opposed to being used only when exploring data analysis options 
(that is, data analysis analysis).  These two works \cite{gair glampedakis, babak et al} will likely soon be merged
in order to produce generic kludge waveforms which represent the entire inspiral.

Another recent set of interesting kludge waveforms are those produced by test particles moving on equatorial orbits of 
a ``quasi-Kerr'' background spacetime (a spacetime which is identical to Kerr, but which has a perturbed quadrupole 
moment, see Ref.~\cite{glampedakis babak} for further details).  Those authors found that, by varying the orbital parameters, one could 
generally find a background-orbit pair (Kerr, geodesic orbit) which resulted in waveforms with a high overlap ($> 90 \%$) 
compared to some pair (quasi-Kerr, geodesic orbit).  
If this holds true in general, the common wisdom that the waveform is a unique signature of the black hole will be cast into doubt.  
It seems likely that this ``confusion problem'' won't persist when one includes the effects of radiation and generic orbits.  Radiation 
may shift the orbit-pairs away from the overlaping regime.  Also, the high overlap required common orbital frequencies, but it is not yet clear 
that generic quasi-Kerr orbits have orbital frequencies (as opposed to continuous spectra).  
Nevertheless, the question of whether or not the confusion problem is general remains open.

\section{Data analysis}

Even if all three classes of waveforms (Capra, Teukolsky, and kludge) were readily available, 
the problem of extracting EMRI waveforms from data would remain non-trivial.
In this section I briefly describe the status of efforts to develop EMRI data analysis algorithms for LISA.  
As promised, this section will be less detailed than the previous section on waveform 
calculations.  A more thorough review of these issues can be found in Ref.~\cite{gair et al}.

Gair et al \cite{gair et al} estimate that a ``brute force'' matched filter search algorithm (akin to LIGO searches for neutron 
star binaries \cite{ligo search}) that uses year-long template waveforms to find EMRIs would require on the order of $10^{40}$ templates, 
rendering it computationally impractical.  Their favored practical alternative is a heirarchical search which 
begins with short duration templates (lasting a few weeks) of modest accuracy.  It should be emphasized that the full year or so worth of data is used 
at this stage---it is only the waveforms which are short.  One still needs a scheme for jumping from one short waveform to the next, but that scheme does not
demand modeling the phase evolution for times longer than a few weeks.
This segment of the search can be thought of as the ``detection stage''
in which EMRI candidates are identified, and relatively modest estimates of their parameters are made.
The search would then proceed toward a ``measurement stage''  which focuses in on narrow regions of parameter space using increasingly accurate and longer-lasting templates, 
and which realizes LISA's full sensitivity (e.g.~measuring $\mu$, $M$, and $a$ with fractional 
accuracy $\sim 10^{-4}$ \cite{barack cutler}).  
They estimate that such a scheme would yield anywhere from tens to thousands of EMRI observations over LISA's lifetime \footnote{This 
estimate neglects sources beyond a redshift $z = 1$.  Although LISA could detect more distant EMRIs, population estimates at such distances 
are unreliable \cite{gair et al}.}.

\begin{figure}
\begin{center}
\epsfig{file=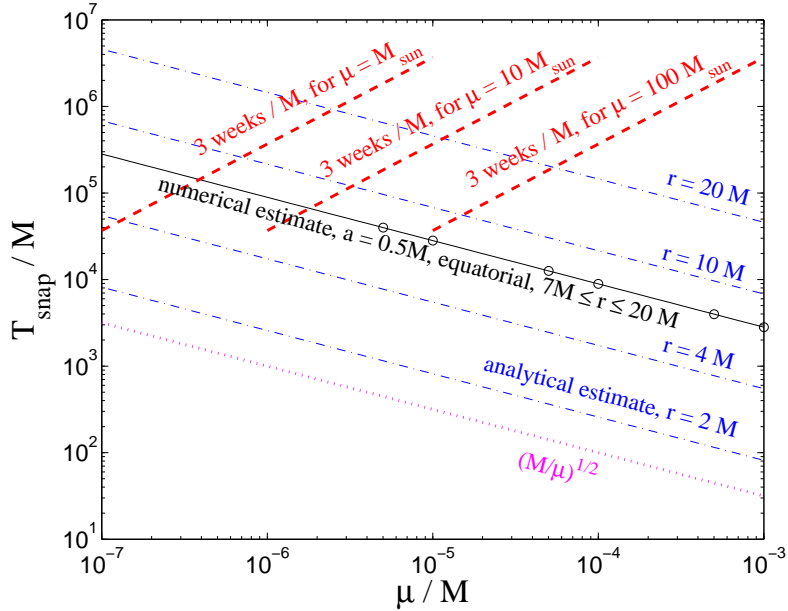,angle=0,width=10.5cm}
\caption{Snapshot timescale $T_{\rm{snap}}/M$ versus mass ratio $\mu/M$ for a black hole spin of $a = 0.5M$, and a geodesic with $e=0.5$, $p=10$, and $\iota = 0^\circ$.  
The (black) solid-line and circles are numerical estimates reported by Glampedakis and Babak \cite{glampedakis babak}.  The (blue) dashdot lines are the analytical estimate (\ref{analytical estimate}) evaluated 
at the indicated orbital radii, $r/M =$ 2, 4, 10, and 20.  
The (magenta) dotted line is the estimate from the scaling argument $T_{\rm{snap}} = M\sqrt{M/\mu}$.  
If waveform snapshots are to be sufficient for detection, they must have $T_{\rm{snap}} > $ 3 weeks.  This boundary is shown as (red) dashed lines for a 
few different test masses $\mu$.}
\label{f:Tsnap}
\end{center}
\end{figure}

In the hierarchical search envisioned in Ref.~\cite{gair et al}, Teukolsky waveforms (or perhaps even kludge waveforms) lasting 
up to a few weeks will likely suffice as templates for the initial detection stage, whereas Capra waveforms lasting up to a few years will be needed for the final measurement stage.  
Let's focus now on the detection stage.  The number of waveform snapshots (waveforms produced by a single fixed geodesic) needed to produce sufficiently 
accurate detection waveforms is currently unknown.  For example, it is possible that snapshots alone might suffice as detection templates.
If so, then the tools for EMRI detection are available already \cite{drasco hughes 2006}.  
The following simple scaling argument would seem to suggest that snapshots cannot be used as detection templates.  
Expand the $\phi$ coordinate of the test mass' world line (a quantity interchangeable with the phase of the waveform for purposes of this rough argument) as follows:
\begin{equation} \label{phi expand}
\phi(t) = \phi_0 + {\dot \phi_0} t + \frac{1}{2} {\ddot \phi}_0 t^2 + \cdots \;.
\end{equation}
Here $\phi_0$, ${\dot \phi}_0$, and ${\ddot \phi}_0$ are constants. The first two terms would be exact for a circular-equatorial geodesic, while the third term represents
the effect of radiation.  A waveform constructed from the geodesic terms alone would produce a phase error of order unity after 
a time $T_{\rm{snap}}$, where ${\ddot \phi}_0 T_{\rm{snap}}^2 \sim 1$.
By dimensional analysis, one would expect ${\dot \phi_0} M \sim 1$ and ${\ddot \phi_0}M^2 \sim \mu/M$.  This suggests that the waveform snapshot would be valid for times
shorter than 
\begin{equation} \label{Tsnap}
T_{\rm{snap}} \sim M \sqrt{M/\mu} \;. 
\end{equation}
This equation predicts that, for EMRIs that could be observed with LISA,  a single snapshot would accurately represent the waveform for times
ranging from a few seconds to a day or so---hardly the few weeks demanded \cite{gair et al} of detection templates.  However, this is only a scaling argument, and it 
neglects potentially significant coefficients which must diverge as the Newtonian limit is approached.  Recently, Glampedakis and Babak estimated values 
for $T_{\rm{snap}}$ by comparing snapshots of Teukolsky waveforms to kludge inspiral waveforms in the case of eccentric-equatorial orbits \cite{glampedakis babak}.  They found, 
even for orbits relatively close to the horizon,  that Eq.~\ref{Tsnap} underestimates $T_{\rm{snap}}$ by a significant factor $T_{\rm{snap}} / (M \sqrt{M/\mu}) \sim 100$ (see Fig.~\ref{f:Tsnap}).   
An alternative analytical estimate  (\ref{analytical estimate}), shown in Fig. \ref{f:Tsnap} and derived in the appendix below, gives similar results.  These results suggest that 
existing waveform snapshots will suffice for detection, at least for some subset of EMRIs that could be observed by LISA.

Lastly, I want to address a topic which is relevant to EMRIs, but which is also of general interest for LISA data analysis as a whole.
One of the key differences between LISA and LIGO-like detectors is that LISA is anticipated to 
find an over abundance of sources.  This very attractive feature also poses a potential problem in 
that waveforms from many different sources will overlap each other, complicating the task of 
cleanly identifying any one of the sources.  For example, the waveforms produced by galactic 
compact binaries (GCBs, e.g.~a pair of white dwarfs) are nearly monochromatic, so that a collection of thousands of these waveforms 
resembles a Fourier basis complete enough to fit essentially any smooth function in their frequency band.  
Therefore, an attempt to first subtract thousands of galactic binary waveforms from LISA data before 
searching for other signals is likely to (i) produce erroneous population statistics for the GCBs 
and (ii) doom further searches by washing away any remaining signals.
In order to avoid this problem, LISA data analysis algorithms will simultaneously search for a variety of 
different sources.  Algorithms that detect single waveforms in isolation are an essential ingredient to this process, 
however the isolated techniques will ultimately be merged into one ``global fit'' algorithm.

When it comes to LISA's general need for global fit algorithms, EMRIs are no exception to the rule.  
Their lower frequencies will likely be hidden by a background of unresolved GCBs, and their 
observable band is expected to overlap with resolvable GCBs and mergers of massive black holes \cite{berti 2006}.  
Recent advances toward global fit data analysis techniques can be found in Refs.~\cite{umstatter et al, cornish crowder}.
Though these works do not deal with EMRIs, similar work must eventually do so, and these are good indicators of 
general progress toward global fit analysis for LISA.  In Ref.~\cite{umstatter et al}, a ``reversible jump Markov chain Monte Carlo'' technique 
is used to simulate the detection and measurement of 100 monochromatic signals.  In Ref.~\cite{cornish crowder}, similar techniques are 
implemented for up to 10 GCBs.  See also the two contributions by J. Crowder and by E. D. M. Wickham to this meeting (GWDAW 10).
These works differ more in their numerical methods than in their theoretical foundations.  They each 
represent implementations of Bayesian Inference \cite{loredo}.    

\section{Provocation}

The conference organizers asked for a review that would provoke discussion.  
In case I have failed to provoke anyone so far, this list of questions (and my \emph{guessed} answers
\footnote{These guesses are for provocation purposes only.  They should not be used in applications 
where injury or property damage may result if said guesses turn out to be wrong.}) might:
Could EMRI detections be made using only the existing waveform snapshots as detection templates? 
(This would work for some EMRIs with $\mu/M \lesssim 10^{-5}$.)
Can kludge waveforms be used as detection templates? (Probably, but other waveforms will be needed in order to determine which kludges suffice.)
Is the conservative self-force needed for detection?  (No, but it will be needed for detailed EMRI measurements.)
Are EMRI waveforms really unique signatures of the background spacetime, or is there instead a Kerr versus non-Kerr confusion problem? (Including the effects of radiation 
and generic orbits will show that there is no confusion problem.)
What are the prospects for global fit data analysis techniques? 
(It is difficult to speculate quantitatively on the prospects for this field.  
However, this type of analysis is a hot topic in many fields that are otherwise unrelated to gravitational wave detection \cite{loredo}.  
Due to the rapid growth in the field, as evidenced by the regular additions to lists of relevant papers \cite{mcmc},
it seems wise to develop as many independent global fit methods as possible.) 
Should we look for IMRIs with ground-based detectors?  (Yes, little is know about the sources, and searches can be done immediately.)
Detailed analyses of any of these questions would likely make a significant contribution toward eventual EMRI observations.

\ack
I thank the Center for Gravitational Wave Astronomy, the University of Texas at Brownsville, and the 
conference organization committees for hosting this meeting, and for inviting me to give this review.  
I thank the authors of Ref.~\cite{babak et al} for discussions, and for permission to include the kludged 
waveforms shown in Fig.~\ref{f:BFGGH}.  
I thank Norichika Sago for sharing the hybrid results that went into table \ref{hybrid vs Teuk}.
and I thank Leor Barack, Duncan Brown, Curt Cutler, Teviet Creighton, {\'E}anna Flanagan, Carlos Lousto, Kip Thorne, and Michele Vallisneri for discussions and encouragement.
I thank the Kavli Institute for Theoretical Physics, at the University of California, Santa Barbara, where this research was 
supported in part by the National Science Foundation under Grant No.~PHY99-07949.
This research was carried out in part at the Jet Propulsion Laboratory, California Institute of Technology, 
under a contract with the National Aeronautics and Space Administration and funded through the 
internal Human Resources Development Fund initiative.

\appendix

\section{Analytic estimate of $T_{\rm{snap}}$ \label{appendix}}

I thank {\'E}anna Flanagan for permission to include this paraphrasing of his analytical estimate for the
time $T_{\rm{snap}}$ over which waveform snapshots are valid.

Suppose that an EMRI's true orbital phase $\phi_{\rm{true}}(t)$ can be approximated
as a quadratic in time, as in Eq.~(\ref{phi expand}), while the orbital phase used to compute a waveform snapshot template $\phi_{\rm{temp}}(t)$ is only linear in time, 
as would be the case if the snapshot was produced by a test mass on a circular orbit.
Take the true and approximate waveforms to have the form $h = A \cos[2 \phi(t)]$, where $A$ is constant, and $\phi(t)$ is the corresponding orbital phase
\begin{eqnarray} 
h_{\rm{true}} &=& A \cos (2\phi_{\rm{true}}) = A \cos ( 2\phi_0 + 2{\dot \phi_0} t + {\ddot \phi}_0 t^2 )~ \label{true}, \\
h_{\rm{temp}}  &=& A \cos (2\phi_{\rm{temp}}) = A \cos ( 2\phi_0 + 2{\dot \phi_0} t )~ \label{temp}.
\end{eqnarray}
Also assume for simplicity that the average overlap integral over a time $T$ is just a time integral, from $-T/2$ to $T/2$, 
of the product of two waveforms, divided by $T$.  The average overlap of the waveforms $h_{\rm{true}}$ and $h_{\rm{temp}}$ is then
\begin{equation}
\fl
V_{\rm{avg}}(T) = \frac{A^2}{2T}\int_{-T/2}^{T/2} dt~\cos(2\phi_{\rm{true}} + 2\phi_{\rm{temp}})
         + \frac{A^2}{2T}\int_{-T/2}^{T/2} dt~\cos(2\phi_{\rm{true}} - 2\phi_{\rm{temp}})~.
\end{equation}
The first term in this expression vanishes rapidly, and is presumably unobservable. The second term dominates
since it approaches zero very slowly (typically on time scale that is many orders of magnitude longer than the time 
needed for the first term to vanish).  The normalized overlap (in the sense that it goes to unity for a pair of identical waveforms) 
of these two waveforms is then given by 
\begin{equation}
V_{\rm{norm}}(T) = \frac{1}{T}\int_{-T/2}^{T/2} dt~\cos(2\phi_{\rm{true}} - 2\phi_{\rm{temp}})~.
\end{equation}
Substitution of the phases read off from the waveforms (\ref{true}) and (\ref{temp}) gives
\begin{equation}
V_{\rm{norm}}(T) = \frac{1}{T}\int_{-T/2}^{T/2} dt~\cos({\ddot \phi}_0 t^2)
 = \frac{C[x(T)]}{x(T)}~,
\end{equation}
where 
\begin{equation}
C(x) = \int_0^x dt~\cos(\pi t^2/2)~,
\end{equation}
is the Fresnel cosine integral, and where
\begin{equation}
x(T) = T \sqrt{\frac{{\ddot \phi}_0}{2\pi}}  ~.
\end{equation}
As in Ref.~\cite{glampedakis babak}, define the time at which the approximate waveform 
is invalid $T_{\rm{snap}}$ to be the time at which the overlap $V_{\rm{norm}}$ drops to 95\%.  Solving $C(x)/x=0.95$ gives $x = x_0 = 0.67 \approx 2/3$, 
which gives
\begin{equation}
T_{\rm{snap}}^2 = \frac{8 \pi}{9} {\ddot \phi_0}^{-1} ~.
\end{equation}
Now use the Newtonian formula (from Sec.~III of Ref.~\cite{finn thorne} with $\Omega = {\dot \phi_0}$ and ${\dot \Omega} = {\ddot \phi_0}$),
\begin{equation}
{\ddot \phi}_0 = \frac{96}{5}\mu M^{2/3}{\dot \phi}_0^{11/3} ~,
\end{equation}
and approximate the orbital frequency $\dot \phi_0$ with Kepler's law $r/M = (M {\dot \phi_0})^{-2/3}$, to find
\begin{equation} \label{analytical estimate}
T_{\rm{snap}} = \left[ \sqrt{\frac{5 \pi}{108}} \left( \frac{r}{M} \right)^{11/4} \right] M \sqrt{\frac{M}{\mu}} ~.
\end{equation}
As expected, this improved estimate diverges in the Newtonian limit $r/M \to \infty$.
Over the range of innermost stable circular orbits of rotating black holes $1 \leq r/M \leq 9$, it differs from 
the result of the simple scaling argument by a factor ranging roughly from $38\% $ to $160$.


\section*{References}

\begin{thebibliography}{10}

%
% Refs. for "Overview"
%
\bibitem{glampedakis review}
K. Glampedakis,
\emph{Extreme mass ratio inspirals: LISA's unique probe of black hole gravity},
Class. Quant. Grav., {\bf 22}, S605--S659 (2005).
%
\bibitem{poisson livrev} 
E. Poisson, 
\emph{The Motion of Point Particles in Curved Spacetime},
Living Rev. Relativity {\bf 7}, 6 (2004), http://www.livingreviews.org/lrr-2004-6 .
%
\bibitem{sasaki tagoshi}
M. Sasaki and H. Tagoshi,
\emph{Analytic Black Hole Perturbation Approach to Gravitational Radiation},
Living Rev. Relativity {\bf 6}, 6 (2003),  http://www.livingreviews.org/lrr-2003-6/ .
%
\bibitem{hughes 2000}
S. A. Hughes,
\emph{Evolution of circular, nonequatorial orbits of Kerr black holes due to gravitational-wave emission},
Phys. Rev. D {\bf 61}, 084004 (2000);
Phys. Rev. D {\bf 63}, 049902(E) (2001);
Phys. Rev. D {\bf 65}, 069902(E) (2002);
Phys. Rev. D {\bf 67}, 089901(E) (2003).
%
\bibitem{brown et al}
D. Brown et al,
in preperation.
%
% Refs. for "The EMR in EMRI"
%
\bibitem{mino 2003}
Y. Mino, 
\emph{Perturbative approach to an orbital evolution around a supermassive black hole},
Phys. Rev. D {\bf 67}, 084027 (2003).
%
\bibitem{drasco hughes 2004}
S. Drasco and S.  A. Hughes, 
\emph{Rotating black hole orbit functionals in the frequency domain},
Phys. Rev. D {\bf 69}, 044015 (2004).
%
\bibitem{drasco hughes 2006}
S. Drasco and S. A. Hughes,
\emph{Gravitational wave snapshots of generic extreme mass ratio inspirals},
Phys. Rev. D {\bf 73} (2006).
%
\bibitem{drasco hughes data}
Waveforms and flux data described in Ref.~\cite{drasco hughes 2006}, and for thousands of other
generic orbital configurations, have been made public at http://gmunu.mit.edu/sdrasco/snapshots/~.  The supercomputers used in that 
investigation were provided by funding from JPL Institutional Computing and Information Services and the NASA Directorates 
of Aeronautics Research, Science, Exploration Systems, and Space Operations.
%
% Refs. for "Capra waveforms"
%
\bibitem{misata} 
Y. Mino, M. Sasaki, and T. Tanaka, 
\emph{Gravitational radiation reaction to a particle motion},
Phys. Rev. D {\bf 55}, 3457 (1997).
%
\bibitem{quwa} 
T. C. Quinn and R. M. Wald, 
\emph{Axiomatic approach to electromagnetic and gravitational radiation reaction of particles in curved spacetime},
Phys. Rev. D {\bf 56}, 3381 (1997).
%
\bibitem{pound poisson nickel}
A. Pound, E. Poisson, and B. G. Nickel, 
\emph{Limitations of the adiabatic approximation to the gravitational self-force},
Phys. Rev. D {\bf 72}, 124001 (2005).
%
\bibitem{rosenthal}
E. Rosenthal,
\emph{Construction of the second-order gravitational perturbations produced by a compact object},
Phys. Rev. D {\bf 73}, 044034 (2006).
%
\bibitem{mino 2004}
Y. Mino,
\emph{Self-Force in the Radiation Reaction Formula},
Prog. Theor. Phys. {\bf 113 }, 733 (2005).
%
\bibitem{hinderer flanagan}
T. Hinderer and {\'E}. {\'E}. Flanagan,
in preparation.
%
\bibitem{capra 2005}
The presentations from this meeting can be found online at http://www.sstd.rl.ac.uk/capra .
%
\bibitem{barack lousto}
L. Barack and C. O. Lousto,
\emph{Perturbations of Schwarzschild black holes in the Lorenz gauge: Formulation and numerical implementation},
Phys. Rev. D {\bf 72}, 104026 (2005).
%
\bibitem{barack ori 2000}
L. Barack and A. Ori, 
\emph{Mode sum regularization approach for the self-force in black hole spacetime},
Phys. Rev. D {\bf 61}, 061502 (2000).
%
\bibitem{barack et al}
L. Barack, Y. Mino, H. Nakano, A. Ori, and M Sasaki,
\emph{Calculating the Gravitational Self-Force in Schwarzschild Spacetime},
Phys. Rev. Lett. {\bf 88}, 091101 (2002).
%
\bibitem{barack ori 2003}
L. Barack and A. Ori,
\emph{Gravitational Self-Force on a Particle Orbiting a Kerr Black Hole}, 
Phys. Rev. Lett. {\bf 90}, 111101 (2003).
%
\bibitem{poisson 2004}
E. Poisson, 
\emph{The gravitational self-force},
gr-qc/0410127.
%
% Refs. for "Teukolsky waveforms"
%
\bibitem{teukolsky 1972}
S.  A. Teukolsky, 
\emph{Rotating Black Holes: Separable Wave Equations for Gravitational and Electromagnetic Perturbations},
Phys. Rev. Lett. {\bf 29}, 1114 (1972).
%
\bibitem{teukolsky 1973}
S. A. Teukolsky, 
\emph{Perturbations of a rotating black hole. I. fundamental equations for gravitational, electromagnetic, and neutrino-field perturbations},
Astrophys. J. {\bf 185}, 635 (1973).
%
\bibitem{ryan}
M. P. Ryan Jr., 
\emph{Teukolsky equation and Penrose wave equation},
Phys. Rev. D {\bf 10}, 1736 (1974). 
This reference gives an interesting alternative derivation of the Teukolsky equation, 
``from a second-order wave equation for the Riemann tensor.''
%
\bibitem{chandra}
S. Chandrasekhar, 
\emph{The Mathematical Theory of Black Holes}
(Oxford University Press, New York, 1983).
%
\bibitem{sago et al 2005} 
N. Sago, T. Tanaka, W. Hikida, and H. Nakano,
\emph{Adiabatic radiation reaction to the orbits in Kerr Spacetime},
Prog. Theor. Phys. {\bf 114 }, 509 (2005).
%
\bibitem{sago et al 2006}
N. Sago, T. Tanaka, W. Hikida, H. Nakano, K. Ganz, 
\emph{The adiabatic evolution of orbital parameters in the Kerr spacetime},
Prog. Theor. Phys. {\bf 115}, 873 (2006).
%
\bibitem{drasco flanagan hughes}
S. Drasco, {\'E}. {\'E}. Flanagan, and S. A. Hughes,
\emph{Computing inspirals in Kerr in the adiabatic regime. I. The scalar case},
Class. Quantum Grav. {\bf 22}, S801 (2005).
%
\bibitem{baumgarte shapiro}
T. W. Baumgarte and S. L. Shapiro,
\emph{Numerical relativity and compact binaries},
Phys. Rep. {\bf 376}, 41 (2003).
%
\bibitem{hdff}
S.  A. Hughes, S. Drasco, {\'E}. {\'E}. Flanagan, and J. Franklin, 
\emph{Gravitational radiation reaction and inspiral waveforms in the adiabatic limit}, 
Phys. Rev. Lett. {\bf 94}, 221101 (2005).
%
\bibitem{teukolsky press}
S. A. Teukolsky and W. H. Press, 
\emph{Perturbations of Rotating Black Hole. III. Interaction of the Hole  with Gravitational and Electromagnetic Radiation},
Astrophys. J. {\bf 193}, 443 (1974).
%
\bibitem{galtsov}
D.V. Gal'tsov.
\emph{Radiation reaction in the Kerr gravitational field},
J. Phys A: Math. Gen. {\bf 15}, 3737 (1982).
%
\bibitem{sago drasco}
N. Sago and S. Drasco, in preparation.
%
\bibitem{khanna}
G. Khanna,
\emph{Teukolsky evolution of particle orbits around Kerr black holes in the time domain: Elliptic and inclined orbits},
Phys. Rev. D {\bf 69}, 024016 (2004).
%
\bibitem{sopuerta laguna}
C. F. Sopuerta and P. Laguna,
\emph{ A Finite Element Computation of the Gravitational Radiation emitted by a Point-like object orbiting a Non-rotating Black Hole}
Phys. Rev. D {\bf 73}, 044028 (2006).
%
%
% Refs. for "kludge waveforms"
%
\bibitem{gair glampedakis}
J. R. Gair and K. Glampedakis,
\emph{Improved approximate inspirals of test-bodies into Kerr black holes},
Phys. Rev. D {\bf 73},  064037 (2006).
%
\bibitem{collins hughes} 
N. A. Collins and S. A. Hughes, 
\emph{Towards a formalism for mapping the spacetimes of massive compact objects: Bumpy black holes and their orbits},
Phys. Rev. D {\bf 69}, 124022 (2004).
%
\bibitem{glampedakis babak}
K. Glampedakis and S. Babak,
\emph{Mapping spacetimes with LISA: inspiral of a test-body in a `quasi-Kerr' field},
Class. Quant. Grav. {\bf 23}, 4167 (2006).
%
\bibitem{peters mathews}
\emph{Gravitational Radiation from Point Masses in a Keplerian Orbit}.
P. C. Peters and J. Mathews,
Phys. Rev. {\bf 131}, 435 (1963).
%
\bibitem{barack cutler}
L. Barack and C. Cutler,
\emph{LISA capture sources: approximate waveforms, signal-to-noise ratios, and parameter estimation accuracy}
Phys. Rev. D {\bf 69}, 082005 (2004).
%
\bibitem{carter}
B. Carter,
\emph{Global structure of the Kerr family of gravitational fields},
Phys. Rev. {\bf 174}, 1559 (1968).
%
\bibitem{ghk}
K. Glampedakis, S. A. Hughes, and D. Kennefick,
\emph{Approximating the inspiral of test bodies into Kerr black holes};
Phys. Rev. D {\bf 66}, 064005 (2002).
%
\bibitem{babak et al}
S. Babak, H. Fang, J. R. Gair, K. Glampedakis, and S. A. Hughes,
\emph{``Kludge'' gravitational waveforms for a test-body orbiting a Kerr black hole},
gr-qc/0607007.
%
\bibitem{sago private}
N. Sago, private communication (2005).
%
% Refs. for "data analysis"
%
\bibitem{gair et al}
J. R. Gair et al,
\emph{Event rate estimates for LISA extreme mass ratio capture sources},
Class. Quant. Grav. {\bf 21}, S1595-S1606(2004).
%
\bibitem{ligo search}
LIGO Scientific Collaboration: B. Abbott et al,
\emph{Search for gravitational waves from galactic and extra--galactic binary neutron stars},
Phys. Rev. D {\bf 72}, 082001(2005).
%
\bibitem{berti 2006}
E. Berti,
\emph{LISA observations of massive black hole mergers: event rates and issues in waveform modeling},
astro-ph/0602470.
%
\bibitem{umstatter et al}
R. Umst{\" a}tter, et al,
\emph{Bayesian modeling of source confusion in LISA data},
Phys. Rev. D {\bf 72},  022001 (2005).
%
\bibitem{cornish crowder}
N. J. Cornish and J. Crowder,
\emph{LISA Data Analysis using MCMC methods},
Phys. Rev. D {\bf 72}, 043005 (2005).
%
\bibitem{loredo}
T. J. Loredo, 
\emph{Computational technology for Bayesian inference}, in 
\emph{Astronomical Society of the Pacific Conference Series, San Francisco, 1999}, edited by R. (Dick) Crutcher 
and D. Mehringer, vol 172, p. 297.
%
% Refs. for "Provocation"
%
\bibitem{mcmc}
See the MCMC preprint service http://www.statslab.cam.ac.uk/$\sim$mcmc/pages/latest.html .
%
% Refs. for appendix
%
\bibitem{finn thorne}
L. S. Finn and K. S. Thorne, 
\emph{Gravitational waves from a compact star in a circular, inspiral orbit, in the equatorial plane of a massive, spinning black hole, as observed by {LISA}},
Phys. Rev. D {\bf 62}, 124021 (2000).
%
\end{thebibliography}
\end{document}